\documentclass[]{aa} %
\usepackage{amsmath} 
\usepackage{graphicx}
\usepackage[usenames]{color}
\usepackage{txfonts}
\usepackage{longtable}
\usepackage{hyperref}
\usepackage{natbib}

\DeclareMathOperator\erf{erf}

\newcommand{\nc}{\newcommand}
\nc{\teff}{$T_{\rm{eff}}$\,}
\nc{\logg}{$\log \,(g)$\,}
\nc{\kms}{\,${\rm km\,s}^{-1}$\,}
\nc{\ms}{\,${\rm m\,s}^{-1}$\,}
\nc{\vsini}{$v \sin \,(i)$\,}      
\nc{\vmic}{$v_{\rm{mic}}$\,}
\nc{\vrad}{$V_{\rm{rad}}$\,}
\nc{\fe}{$\rm{[Fe/H]}$\,} 
\nc{\li}{$A(\rm{Li})$\,} 
\nc{\msun}{$M_{\rm{\odot}}$\,}

\begin{document}

\title{Tachoastrometry:  astrometry with radial velocities}


   \author{
        L. Pasquini\inst{1}
        \and
        C. Cort\'es \inst{2,3}
        \and
        M. Lombardi \inst{4}
          \and
          L. Monaco \inst{5}
          \and
          I. C. Le\~ao \inst{6}
          \and
          B. Delabre \inst{1}
          }
          

     \institute{
ESO, Garching bei M\"unchen, Germany
                \and
Departamento de F\'{i}sica, Facultad de Ciencias B\'asicas, Universidad Metropolitana de Ciencias de la Educaci\'on, Av. Jos\'e Pedro Alessandri 774, 7760197, \~Nu\~noa, Santiago, Chile      
                \and
The Millennium Institute of Astrophysics (MAS), Santiago, Chile
                  \and  
University of Milan, Department of Physics, via Celoria 16, 20133
, Milan, Italy
                 \and
                 ESO, Chile
                 \and 
                 Departamento de F\'isica, Universidade Federal do Rio Grande do Norte, Natal, RN, 59072-970 Brazil
                 }
   \date{Received; accepted}

\abstract
{
Spectra  of composite systems (e.g., spectroscopic binaries) contain spatial information that can be 
retrieved by   measuring  the radial velocities (i.e., Doppler shifts) of the components in four observations with the slit rotated  by 90 degrees in the sky. 
}
  { 
We aim at  developing a framework to describe the method  and 
  to test its capabilities  in a  real  case. 
   }
  { By using basic concepts of slit spectroscopy we show  that the geometry of 
   composite systems  can be reliably  retrieved by  measuring only radial velocity differences 
   taken with different slit  angles. The spatial resolution  is  determined by the precision with which  differential radial velocities can be measured. 
   }
   {
 We use the UVES spectrograph at the VLT  to observe the  known spectroscopic binary star HD~188088 
(HIP~97944), which has a  maximum expected separation of 23 milli-arcseconds. We measure an astrometric signal in 
radial velocity of 276~\ms,  which corresponds to a separation between the two components at the time of the 
observations of 18 $\pm2$  milli-arcseconds. The stars were aligned east-west.
We  describe a simple optical device to  simultaneously record  pairs of spectra rotated by 180 degrees, thus reducing systematic effects. 
We compute and provide the function expressing   the shift of the centroid of a seeing-limited image  in the presence of a narrow slit.
}
 { The proposed technique is simple to use and our test shows that it is amenable for deriving astrometry with milli-arcsecond accuracy or better, beyond the diffraction limit of the telescope. The technique  can be further improved by using simple devices to simultaneously record  the spectra with 180 degrees angles. This device together with an optimized data analysis   will further reduce the measurement errors. With tachoastrometry,  radial velocities and  astrometric positions can be measured simultaneously for many double line system binaries in an easy way.  The method is not limited to binary stars, but  can be applied to any astrophysical configuration in which spectral lines are generated by  separate (non-rotational symmetric) regions.    
 }

\titlerunning{Tachoastrometry}
\maketitle
%

\section{Introduction}
The quest to obtain high angular resolution and astrometric information from ground-based telescopes is  quite pressing, 
in particular after   adaptive optics (AO) and optical interferometry have shown its tremendous potential  for 
obtaining sharp images and precise astrometry.
At the same time, prompted mostly by planet search and the study of 
stellar oscillations, radial velocity  precision has dramatically improved as well \citep{Anderson_etal2008}, 
 reaching below the \ms \citep{Mayor_etal2003}. 
The basic idea behind this work is that astrometry and radial velocity are in many instances intimately 
connected. 

Three decades  ago a method to determine the distance between 
two stars of a binary system using radial velocity measurements was suggested \citep{Beckers1984}. 
In this paper we reconsider this idea,  prove its feasibility,  discuss its application  to different astrophysical situations, and provide a simple optical concept to  improve the accuracy of the method.  
In the last 15 years, several spectroscopic techniques have been developed to 
retrieve  astrometric information. Following the seminal paper by   \citet{Beckers1983},   
\citet{Bailey1998}  developed spectroastrometry and applied it to the observation of young stars. 
Spectroastrometry is now used with success on other objects,  from planetary nebulae \citep{BlancoCardenasetal2014}  to AGN nuclei \citep{Gneruccietal2013}.  The technique proposed in this work is related to spectroastrometry, 
but it is at the same time substantially different because it ignores the spatial profile of the 
recorded spectra, and relies uniquely on measured Doppler velocities (which is why the name 
tachoastrometry is used). 

With tachoastrometry the measurable angular separation 
depends  {\it \emph{only}} on its abilty to measure Doppler shift differences  
in the spectral lines  of  the observed objects. The achievable angular resolution depends  
{\it \emph{only}} on  the ratio between the  precision of the Doppler measurement and the slit scale.  
 
\section{  Concept}

The fundamental concept is very simple. In a grating spectrograph, the image at the telescope focal plane 
(or the slit) is re-imaged and dispersed on the spectrograph detector. A source 
moving in the slit plane will move in the detector plane, by an amount  given by \footnote{ We neglect in this equation important effects, such as grating anamorphism. In a real system $\Delta x$  changes with  wavelength. }  

\begin{equation}
 \Delta x_\mathrm{det} = \Delta x_\mathrm{slit}  \frac{F_\mathrm{cam}}{F_\mathrm{col}}
,\end{equation}where $\Delta x_\mathrm{det}$ and $\Delta x_\mathrm{slit} $ are the shift at the slit and at the detector, and $F_\mathrm{cam}$ and $F_\mathrm{col}$ are the focal length of the camera and of the collimator, respectively. 

By moving the object along the slit (i.e., perpendicular to dispersion), the spectrum is shifted along the spatial direction; 
by  moving the object across the slit in the direction of the dispersion, a wavelength shift is thus produced in the observed spectrum, which will induce an observed Doppler shift.  The amount of the shift is  given by (1),  but can be transformed in \kms by using the spectrograph linear dispersion ($\Delta\lambda / \mathrm{mm}$)  and the wavelength:  
\begin{equation}
\Delta\lambda = \Delta x_\mathrm{det} \times \frac{\Delta\lambda} {\mathrm{mm }}  
.\end{equation}Using the standard $ \Delta \mathrm{v} = c \times \frac{\Delta\lambda}{\lambda} $, this brings to 

\begin{equation} 
\Delta \mathrm{v} = \mathrm{c} \times \Delta x_\mathrm{slit} \frac{F_\mathrm{cam}}{F_\mathrm{col}} \times \frac{\Delta\lambda / \mathrm{mm} }{\lambda} 
.\end{equation}

This can also be expressed  in arcseconds  for any spectrograph by transforming $\Delta x_{slit}$ into $\alpha$ (arcseconds). 
A simple, convenient  way of expressing the shift is given by 

\begin{equation}
\Delta \mathrm{v}({\rm km\,s}^{-1}) = \alpha (arcsec)\times \frac{\mathrm{c}}{\mathrm{R}} 
,\end{equation}where  $\mathrm{R}$ is the spectrograph resolving power  for $1~\rm{arcsec}$ slit aperture and $c$ is the speed of light in \kms. 
In a conventional  spectrograph the resolving power  changes with wavelength, and so will the scale conversion factor. In an echelle spectrograph, R is approximately  constant
with wavelength, and so is the conversion factor. 

These are optical relationships. In real observations of a  point-like, seeing-limited  astronomical object,  the 
 shift of the image after the spectrograph  slit  is smaller than  the shift of the object in the focal plane 
 because the effects of a finite slit must be taken into account.  Its computation is given in the Appendix. 
In all cases,  artificial Doppler shifts induced by the slit centering error  may be  considerable and this is a well-known limitation to  measuring precise radial velocities. 
A considerable effort has been devoted to eliminating this effect, either  by  scrambling the source light with fibers and  other optical systems,  
as done, for instance in HARPS (Mayor et al. 2003),  or by using a  gas cell in the optical path as a direct reference source  \citep{Beckers1977}. 

Calling $RV_m$ the measured radial velocity of a star (formally   called measured Doppler shift;  see  
\citet{LindegrenDravins2003} for a definition  of radial velocity, but hereafter we use  radial velocity as a common term), this can be expressed as
 
\begin{equation}
 {RV}_m = {RV}_{true} + {RV}_{spectro} + {RV}_{slit}
 \label{eq:radialV}
\end{equation}
 
\noindent where ${RV}_{spectro}$  and ${RV}_{slit}$ are the shifts induced by  spectrograph instabilities  and by object centering errors, respectively. 
 
The aim of this work is to show that in special circumstances, for instance in the presence of a binary star, instead of trying to eliminate  the
contribution of ${RV}_{slit}$, this  can be successfully used to retrieve the geometry of the system.

For a spectroscopic binary with double line system (SB2) the measured RV difference between the two components will not be determined only by the
true RV difference between the stars, but  will also contain a geometrical component produced by the 
distance between the two stars along the spectrograph dispersion axis. 
Because the two stars of the spectroscopic binary are observed simultaneously and along the same optical path,  ${RV}_{spectro}$ is the same for 
both stars, so it will not contribute to their RV difference. 

\begin{figure}[h]
\resizebox{\hsize}{!}{\includegraphics{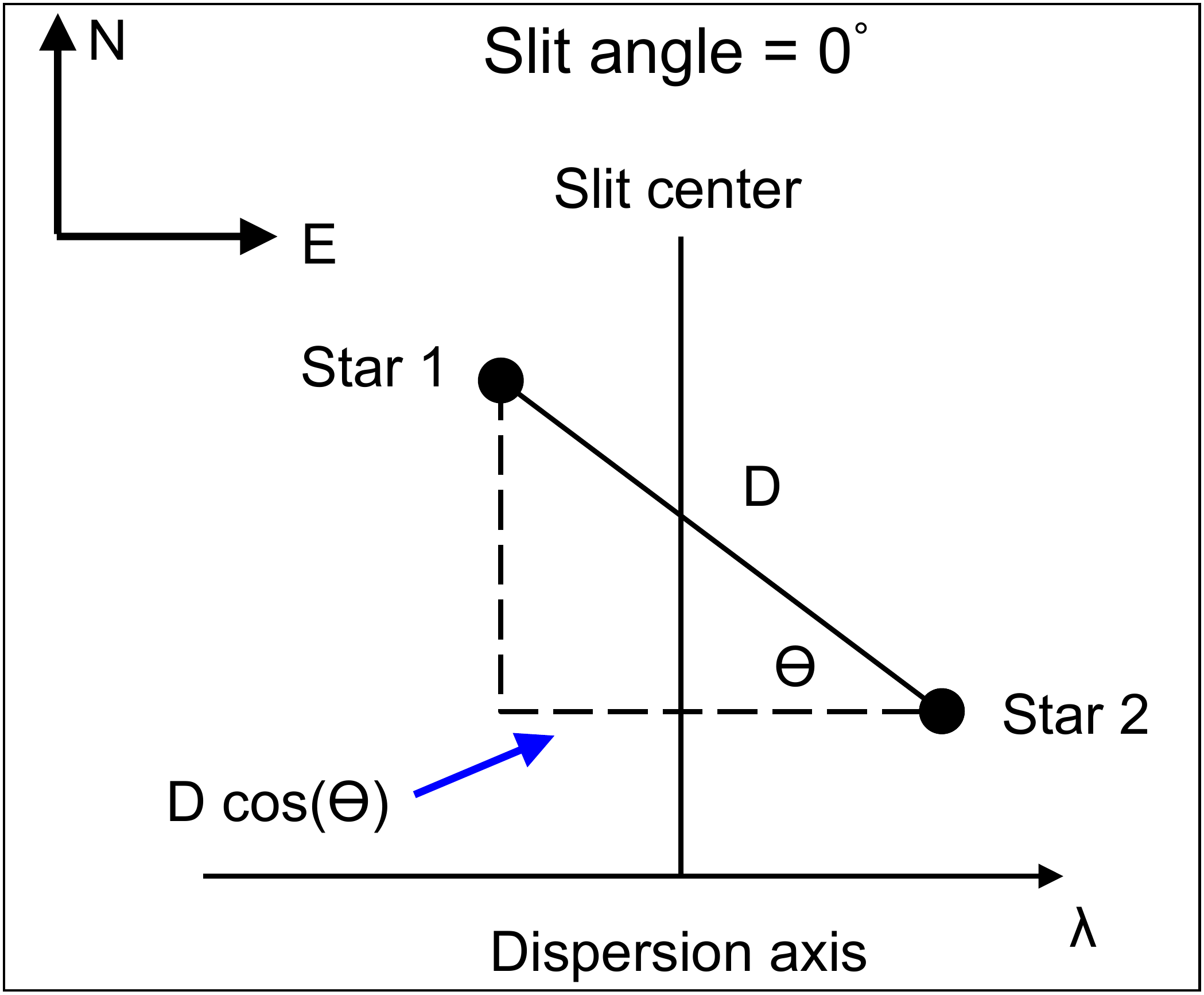}}
\caption{Description of the geometry of the stars and spectrograph slit during the first observation: the observed difference in radial velocity between the stars  depends on their true difference in  RV and  on their separation in the sky. }
\label{fig1}
\end{figure}

The principle  is shown in Fig. \ref{fig1}, with two stars separated by a distance $D $ in the sky,
and a distance $ D\cos \theta$ projected along the  slit width, with $\theta$ being the projected angle between 
the line joining  the two stars and the dispersion direction. In the example we assume that the  two stars have true radial velocities  $RV_1$ and $RV_2$ at the source. 

\begin{figure}[h]
\resizebox{\hsize}{!}{\includegraphics{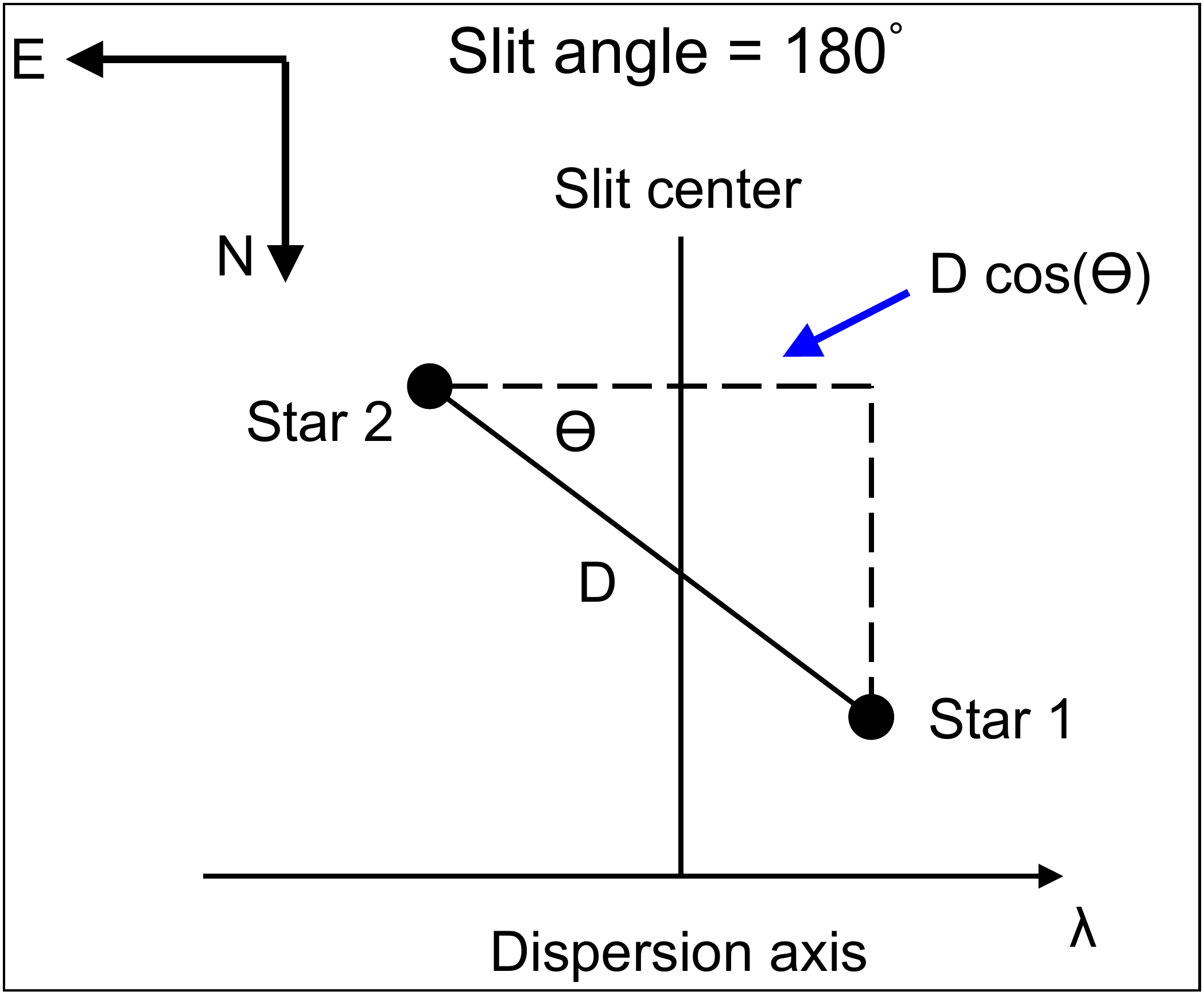}}
\caption{As in Figure~\ref{fig1}, but now the slit has been rotated by $180$ degrees. The measured difference in radial velocity between Figs.~\ref{fig1} and~\ref{fig2} will depend on the projected separation of the stars along the dispersion.}
\label{fig2}
\end{figure}

With the  first observation, we measure therefore a RV difference  $\Delta V_1$ (see Figure~\ref{fig1}):  
\begin{equation}
\Delta V_1 = RV_1 - RV_2 - D\cos \theta  
\label{eq:DV1}
\end{equation}
A  second measurement is acquired  by rotating the slit by $180$ degrees in a  time sequence short enough that 
$RV_1$ and $RV_2$ can be considered constant;  the geometry is given in Figure~\ref{fig2}.  In this case, the difference in the observed RV of the stars 
 is given by 
 
\begin{equation}
\Delta V_2  = RV_1 - RV_2 + D\cos \theta 
\label{eq:DV2}
\end{equation}

\noindent therefore, with a  simple difference we find 

\begin{equation}
2D\cos \theta =  \Delta V_2 - \Delta V_1 
\label{eq:Diff1}
\end{equation}
 
 By taking two additional exposures,  rotating the slit orientation by $90$ and $270$ degrees    and measuring $\Delta V_3$ and $\Delta V_4$, 
  the full geometry of the system is  solved:
 
\begin{center}
\begin{equation}
 \tan \theta =\frac{\Delta V_4-\Delta V_3}{\Delta V_2- \Delta V_1} 
\label{eq:theta}
\end{equation}  
\end{center}

\begin{equation}
\centering
D=\frac{\sqrt{(\Delta V_4-\Delta V_3)^2 + ( \Delta V_2- \Delta V_1)^2}}{2}
\label{eq:D}
\end{equation} 
 
The astrometric information (separation and angle) is obtained  by measuring 
differences in radial velocities;  therefore, the  sensitivity of the method is  determined by the capability of 
measuring precise radial velocities. 
This is extremely attractive  for  cool stars, for instance, that have thousands of narrow spectral lines, for which a
RV precision  close to or  better than $1$~\ms  can be obtained.  Because we measure velocity differences 
in the same spectrum, and in spectra taken within a short time interval, 
most systematic effects are canceled out and there is no need for a long-term instrumental stability.
To obtain  a reference example, a radial velocity difference  of
$1$~\ms on UVES at the VLT \citep{Dekker_etal2000},  for
which $1~\mathrm{arcsec}$  is equal to $7\!.5$~\kms,  
corresponds  to an angular separation   $\alpha =  1\!.33\times10^{-4}~\mathrm{arcsec}$ (133 micro arcseconds), 
which is  several times better than the VLT diffraction limit in the visible. 
If not limited by other factors, this method will therefore open  interesting opportunities.
 
\section{Measurement uncertainties }
 
\subsection{ {\bf Instrumental} error budget}

 Relations 9 and 10 show that the  astrometric precision  depends on the error on how 
 the RV difference can be measured  (hereafter $\delta DV$) as 
 
 $$\delta \tan \theta / \tan \theta =   2 \delta DV / DV $$
 
  and  
  
$$  \delta D  =    \frac {\delta DV}{\sqrt{2}},    $$
provided, of course, that distance and velocity are expressed in the same units.  
As predicted, the precision with which 
 the radial velocity difference  can be measured is directly related to the precision of the geometrical parameters.
 This error assumes that other source of uncertainties, such as the rotation angle of the slit, are negligible. This is the case for 
 all modern instrument adaptors. Another source of instrumental uncertainty is  the stability with time of the linear dispersion $\Delta\lambda$/mm. This can vary, for instance, because of variations in temperature or pressure. UVES  is a very stable instrument, so linear dispersion variability is not an issue within the duration of  one cycle and even for much longer timescales. In our tests we   checked the spectrograph pressure and temperature stability  during our observations; for less stable spectrographs, frequent wavelength calibrations should be taken. We note that it is possible to use calibrations taken with a small slit and use them for the wide slit observations  because what matters is  the stability of the linear dispersion, which does not depend on the slit width. 
  
 \subsection{Seeing,  slit losses}
 
In all cases of astrophysical interest the two sources will appear blended  because of the diffraction limit of the telescope 
and because of the  seeing. 
Seeing will not  directly affect the measurements because at these scales (much less than one arc-second) isoplanatism is guaranteed in  all atmospheric 
conditions even at  visible light;  this implies that both stars
images suffer  exactly of the same wavefront deformations induced by seeing. The images of the stars will 
wander on the focal plane, but always in the same way  for 
both objects, which is important for our case. This implies that the relative position between the two stars will not change, and therefore the difference in radial velocities induced by their geometric distance will not be affected by the seeing. 

When a small slit is used, because the stars are not perfectly centered in the slit, their  flux is cut asymmetrically by the slit jaws.  
The   vignetting will introduce a systematic effect, as discussed in the introduction, that will affect the RV of both stars in a similar  
way. However, the two stars do not have exactly the same position.  In the presence of a narrow slit,  the distance measured after the slit is smaller than the real distance in the sky. This effect will be small in most circumstances, because the stars will not be  very far apart, but still   present and so a correction must be applied.  The measured distance between the stars  will be compressed with respect to the real distance in the sky by a quantity given by equation \ref{bary}. Our tests include observations with narrow and wide slit and the results are in excellent agreement with what predicted by  equation  \ref{bary}. 

In order to safely cope with this problem we will   use a large slit.  This may induce large variations in the 
observed spectral profiles between the four observations if seeing variations between the  exposures occur.  
With a large slit  the spectrograph point spread function (PSF) is determined by the seeing. 
Large PSF variations may complicate the analysis and the measurement of the RV,  because the  observations separated by 180 degrees are not carried out 
simultaneously. In general, variation of the line profiles  will   lower the precision of the RV measurements.  
The appropriate measurement of the RV and a composite and variable instrumental line profile  might introduce noise in  the measurements. The problem of fitting a double function without knowing the exact PSF has been addressed in the literature, and, in order  to derive accurate radial velocity measurements in composite spectra of binary stars, ad hoc methods have been  developed.  For instance \citet{Zuckeretal2003} improve the performances of a simple cross-correlation analysis  by optimizing the spectra of each component with a proper spectral match. We consider the optimization of the data analysis to be beyond the scope of the present work, but we believe   that the method can be improved by  simultaneously acquiring  observations at different angles 
by constructing a simple optical device, and we present a possible design for such a device in Sect. 7. 


\subsection{Source variability}

Tachoastrometry  assumes that the sources have constant radial velocity during the observations. The validity of this   assumption depends on the nature of the object observed  and on the precision requested. The best test cases with which to investigate the  validity of this hypothesis  are solar-type stars. 
In these stars stellar oscillations, long-term magnetic cycles, and rotational modulation affect the radial velocity constancy with different timescales. 
All these effects must be taken into consideration when searching for low mass planetary companions, as in the case of $\alpha$Cen B, which was studied by \citet{Dumusque_etal2012}. 
Magnetic cycles, similar to the solar 11-year cycle, can induce  radial velocity variations of several \ms \citep{Dumusque_etal2012}, but these cycles are effective on long timescales of several years, which are generally of no interest for tachoastrometry. Instead, solar-type oscillations  affect radial velocities with short timescales, at the level of  a few \ms \citet{Kjeldsen_etal2005}, with an amplitude that depends on the stellar effective temperature and gravity. Similarly, rotational modulation can produce radial velocity modulation at a level of up to several \ms depending on the chromospheric activity level of the stars \citep{Dumusque_etal2012, Paulson_etal2004}. 
For astrometric measurements that aim at the highest precision comparable to or below the \ms, these effects must be taken into account. 
As far as stellar oscillations are concerned, many short observations can be acquired, or observations long enough (on the order of 15  minutes) to average out the oscillation signal. Rotational modulation  should instead be modeled, as done in exoplanet searches. It is  likely, however,  that the presence of this variability will set the 
ultimate precision of tachoastrometry for  binary stars to a few \ms. 
 
\section{ The test} 

We tested the method by observing the binary  HD~188088 (HIP~97944) with the UVES spectrograph on the VLT  \citep{Dekker_etal2000} with the standard red arm setup centered at 580 nm. 
The star was carefully selected because the most appopriate object has to fulfill a number of characteristics: 
the expected effect should be conveniently measurable, therefore the angular  separation should vary between a few tenths and a few hundreds of arcseconds; 
the star should be a double line spectroscopic binary, to see  both line systems;  the stars should also be  late-type and slow rotators, in order to 
obtain  precise radial velocity measurements. After all the criteria were applied, we chose HD188088 (HIP97944), whose characteristics are summarized in 
Table~\ref{tab:param}. 
HD~188088 is a binary formed  by 
 two K3V stars \citep{Torres_etal2006}, with similar masses and luminosities ($\sim\!0.86~M_{\rm{\odot}}$) with a period of  $46.817~d$ and a high eccentric ($0.69$)
 orbit  \citep{Fekel_Beavers1983}.

\begin{table}[!h]
\centering
\caption{Orbital elements of HD~188088}
\begin{tabular}{lrl}
\hline \hline
$P$&$46.817~d$& \citealt{Fekel_Beavers1983} \\
$e$&$0.69$& \citealt{Fekel_Beavers1983} \\
$K_1$&$48.8$\kms& \citealt{Fekel_Beavers1983}\\
$K_2$&$48.8$\kms& \citealt{Fekel_Beavers1983}\\
$M_1\sin i$& $0.85$\msun & \citealt{Fekel_Beavers1983}\\
$M_2 \sin i$& $0.86$\msun & \citealt{Fekel_Beavers1983}\\ 
ST& K3V,K3V& \citealt{Torres_etal2006}\\
$ v\sin i$  & 2.0 \kms& \citealt{Torres_etal2006}\\
$\pi$ &$71.18\pm0.42~\mathrm{mas}$&\citealt{van_Leeuwen2007}\\
$a_1\,\sin i$ & $2.27\mathrm{10^{7}~km}$& \citealt{Pourbaix_etal2004}\\
$a_2\,\sin i$ & $2.26\mathrm{10^{7}~km}$& \citealt{Pourbaix_etal2004}\\
\hline
\end{tabular}
 
\label{tab:param}
\end{table}

\begin{figure}[h]
\includegraphics[scale=0.40]{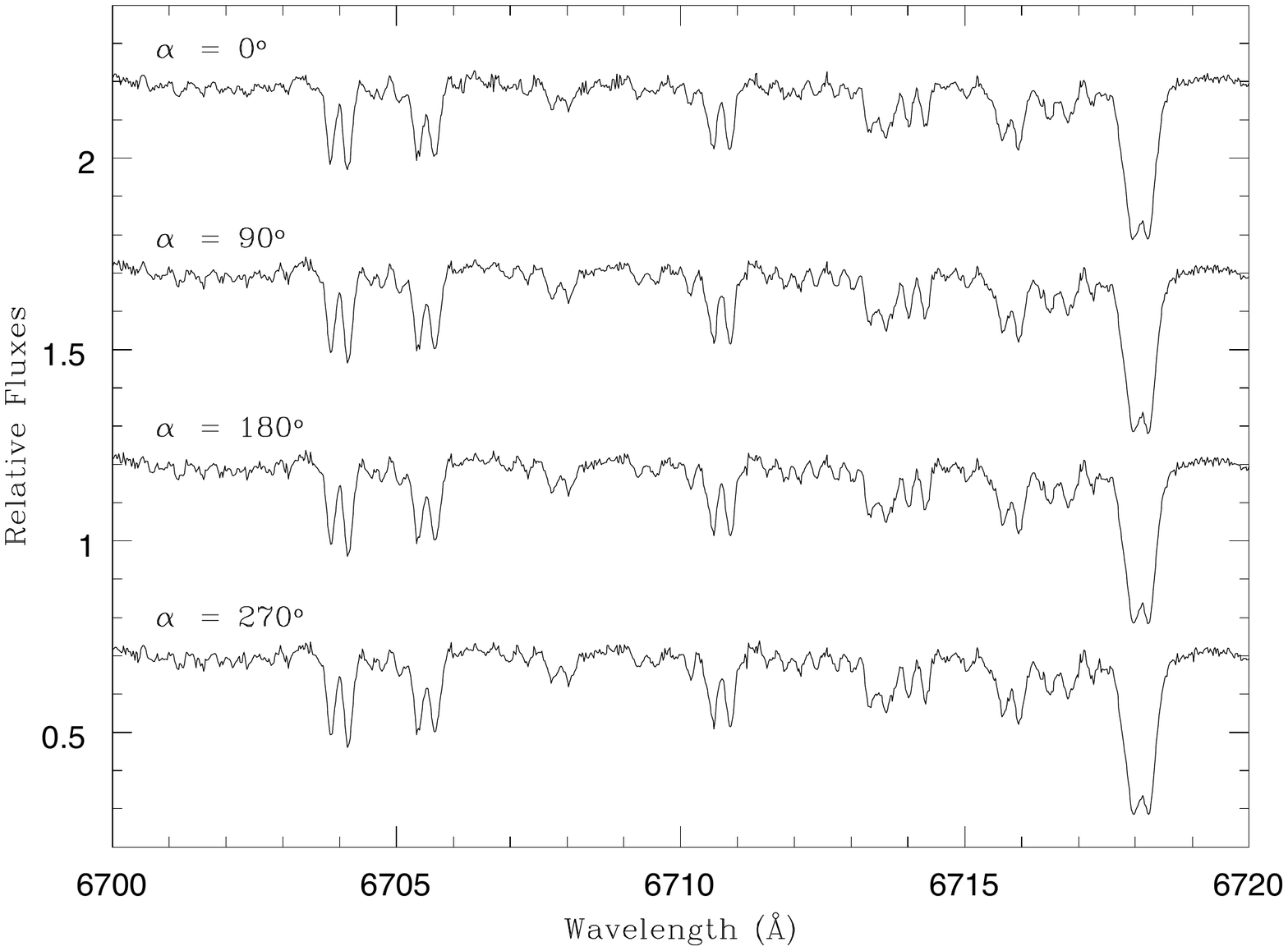}
\caption{The spectra HD~188088 for different slit angles. The lines of the two stars are clearly  separated in  the spectra.}
\label{fig3}
\end{figure}

The observations were carried out as technical tests on September 6, 2013.  Two  cycles were obtained: the first  using a slit width of 1 arcsecond;  single integrations were short, thanks to the  relative brightness of the sources. For the second cycle we used a wide (5 arcseconds) slit width.  By acquiring these two cycles  we are able to the  compare the results with the predictions of the effects expected by  using a small slit. 
Each cycle is composed of seven observations, because observations at 0 degree orientation were repeated, even after a full 360 degree rotation. 
 Each cycle took in total less than 13 min, each exposure being $40$ or $60~$s long, providing a signal-to-noise ratio ($S/N$) of  at least 250  at $5750~\AA$ for each spectrum. The seeing  during the observations was good, and almost constant around 0.7 arcseconds, and the values are reported,  together with  the other  main observation parameters  in Table~\ref{tab:obs}.  Some reduced spectra around the LI ($\sim$6708 $\AA$) are presented in Fig.~\ref{fig3}, where the duplicity of the spectra is evident, and the lines of the two stars are clearly separated.    

\begin{table}[!h]
\centering
\scriptsize
\caption{Ephemerides of the observations of HD~188088, time is UT on September 6, 2013. }
\begin{tabular}{lcrrcc}
\hline \hline
&Slit angle&    $S/N$&  Seeing& Observation&Exposure Time\\
&$(^{\circ})$& & & Hour& (s)\\  
\hline
\smallskip
First  &$0$ &     $     326 $ & $       0.72$ & 03 39 & $60$\\
          & $0$ &    $       262 $ & $    0.7 $ &    03 41 &  $40$ \\
Cycle&$90$ & $  260 $ & $       0.73$ & 03 45 & $40$\\
          &$180$ & $    261 $ & $       0.73$ & 03 49 & $40$\\
  Slit         &$270$ & $       264 $ & $       0.65$ & 03 52  & $40$\\
     1"      &$360$ & $ 261 $ & $       0.65$ & 03 56 & $40$\\
              & $0$ &    $       255 $ & $    0.7 $ &    03 58  &  $40$ \\
\hline
\smallskip
Second & $0$ & $        247 $ & $       0.77$ & 04 01    & $30$\\
              & $0$ & $ 286 $ & $       0.77$ & 04 03   & $40$\\
Cycle  &$90$ & $        286 $ & $       0.72$ & 04 06   & $40$\\   
              &$180 $ & $       287  $ & $ 0.71$ &      04 10   & $40$\\
  Slit        &$270 $ & $       286  $ & $ 0.72$ &      04 13    & $40$\\
  5"       &$360 $ & $  288  $ & $      0.72$ & 04 17    & $40$\\
           & $0$     & $        288 $ & $       0.77$ & 04 20   & $40$\\
\hline
\end{tabular}
 
\label{tab:obs}
\end{table}
\normalsize


\section{Results}

After data reduction, the radial velocity analysis was carried out by using a digital cross-correlation mask \citep{melo_etal2001}.  
We used  two Gaussian functions  to fit the double peaked cross-correlation function (CCF) \citep{Tonry_Davis1979}  and to measure the RVs of each 
star. Since the rotational velocity of both components is low (see Table~\ref{tab:obs}), the Gaussian function 
is adequate to obtain a good representation of the CCF profiles, and we leave  the  Gaussian  widths, intensities, and centers as free parameters.  The results of the single measurements are given in  Table~\ref{tab:RVS}  in which the widths of the Gaussian function $\sigma$ resulting from   the best fit are also reported. 
Figure 4 shows a typical CCF profile, where it can be noticed that  the CCF has low level side lobes that alter the continuum, most likely produced by an imperfect match of the mask used with the stellar spectrum, or by a low level residual fringing.  For future observations, it might be useful to observe a slowly rotating single star of similar spectral type  to produce an empirical CCF profile, or to optimize the digital mask to the spectrum of the observed star.   
 We have performed many tests   by varying the CCF  fit window over a large range. A small window allows very good fits of the CCF cores (and therefore a very small nominal error in the centers), but poses poor constraints on the  wings, leading to unacceptable differences in the separations between the CCF peaks in the two cycles. Large windows are very sensitive to the continuum fluctuations, fail to  satisfactorily reproduce the cores, and thus provide a large error.  However,  while the  RV measurements can vary by several tens of \ms depending on the fitting window  adopted, it is remarkable that the effects are systematic and the $\Delta V_i - \Delta V_{i-1} $  measurements are extremely robust, within a few ($\pm 5$) \ms independent of the continuum window chosen.  
 In Table \ref{tab:results} the measured velocity differences and results are summarized. In the following we use the results from a $\pm 50$\kms window. 

Cycle 1: The error estimated by the fit for each radial velocity is 19\ms, constant for all observations, and the same for the two stars. The repeated observations at 0 and 360 degree show a good  consistency in the radial velocity difference between the two stars. The average difference for this angle is of 13.376 \kms, with a maximum deviation from the average  of 26\ms. We use this average value  for  $\Delta V_1$ for 0 degrees for cycle 1. 

We note that the 0-360 degree observations made at the end of both cycles are slightly smaller than the one made at the beginning. This is consistent with the fact that, according to the published ephemerides,  at the time of the observations the phase was just above 0.5, and the radial velocities of the two stars are predicted to  become closer with time.  

We believe that the difference in the measured values  is the most realistic  error estimate for the velocity difference $\Delta V$ in the CCF.  It is worth noticing that  the last observation of the cycle was performed after a full 360 degree rotation of the slit. The results show that no major effects are produced by the full adapter rotation, because the last two exposures of the cycle  agree at better than 10 \ms. The difference between the 90 and 270 degree observations is negligible $\Delta V_4  - \Delta V_3$=(0 \ms) , while the difference between  0 and 180 degrees, $\Delta V_2 - \Delta V_1 $, is quite pronounced (185 \ms), showing that at the time of the observations the two stars were almost perfectly E-W oriented.

Cycle 2: The estimated error on the single measurement is  higher (32 \ms) and the $\sigma$ slightly larger. This is expected from the use of the large slit.
The measurement at 0 and 360 degrees, are also  consistent, with an average value of $\Delta V_1$=13.300 \ms and a maximum deviation from the average of 17 \ms . Also in this case, the  largest difference is between the 0 and 180 degree orientation, with $\Delta V_2 - \Delta V_1 $ =  276 \ms, while 
$\Delta V_4  - \Delta V_3$ = -4\ms  confirming the negligible separation of the two stars in the N-S direction. 

\begin{figure*}[h]
\includegraphics[scale=0.45]{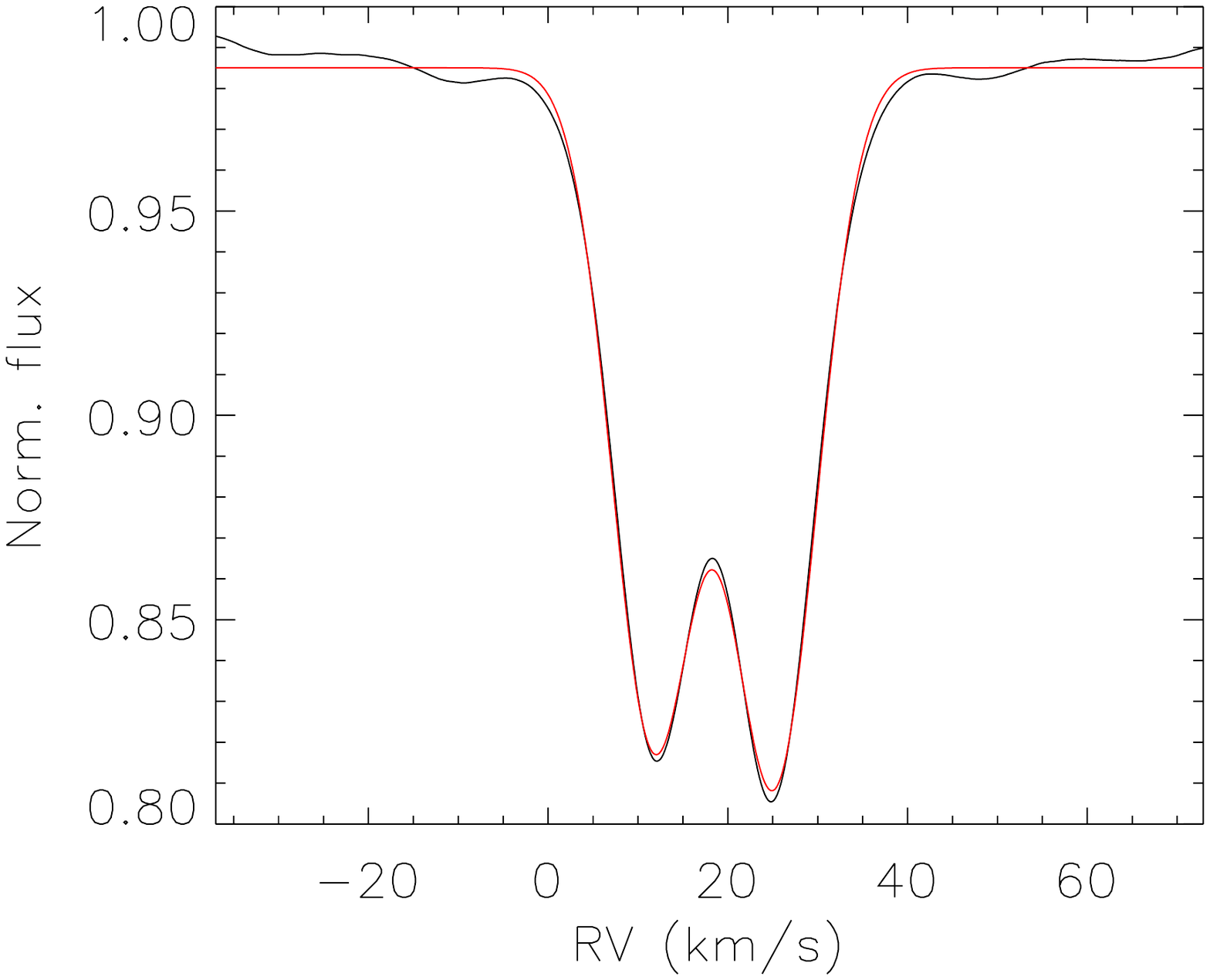}
\caption{Example of cross-correlation function (black line) and fit (red line) for our observations.}
\label{ccf}
\end{figure*}

Both cycles are very consistent. They show a pronounced $\Delta V_2 - \Delta V_1 $ difference, while 
the $\Delta V_4  - \Delta V_3$ difference is negligible (0 and -4\ms respectively for cycle 1 and 2) . 
This indicates that at the moment of the observations the two components of the binary were
almost perfectly aligned E-W. The resulting  distance is of 12 $\pm 2$ milliarcseconds for cycle 1 and of 18 $\pm 2$ milliarcseconds  for cycle 2, the error being 
determined by  the error formula and a $\delta$DV of 26 \ms, as determined from the repeated observations at 0 and 360 degrees for both cycles.
As predicted in section 3.2, the cycle1 observations provide a smaller separation, because they were acquired with a small slit. 

We  estimate  the correction to be applied to the first cycle expected by the use of 1 arcsecond slit. We used as input to equation \ref{bary}  the observational values (seeing=0.7 arcsec, separation between the stars of 0.018 arcseconds, as measured in the second cycle, slit width 1 arcsec).  
 Equation \ref{bary} predicts  for the first cycle an observed value of   $\Delta x_{obs}$ = 11 milliarcseconds, and we measured 12. 
  A correction of 7 milliarcsecond has to be added to the cycle 1 result.  
 This brings the corrected Cycle 1 distance to 19 milliarcseconds, in excellent agreement with the wide slit cycle 2 results. 
 The correction to be applied for the cycle 1 observations (seeing=0.7 arcseconds, slit width =1 arcsecond) is therefore not negligible and critically depends on the seeing value. For the UVES observations, we have two independent  ways of measuring the seeing: from the height of the observed spectra, and from the values provided by the telescope at the beginning and at the end of the observations. They agree very well and, as seen from Table 2, the seeing varied by less than $\pm 5 \%$ during our observations. In order to evaluate the error associated with the correction applied, we use the simplified formula (12). By differentiating equation (12) 
 the uncertainty for the correction is found: $\delta \langle x \rangle / \langle x \rangle = 2 \delta \sigma/\sigma$. This implies that the uncertainty associated with the 
correction applied to cycle 1 is less than $10 \%$ of the correction, or 0.7 milliarcseconds.  This shows a robust result but, as we anticipated, observing with a slit wide enough to avoid slit corrections seems  the safest choice.  It is nevertheless very encouraging that  the two cycles give results in excellent agreement, providing at the same time a nice test of the whole procedure. 

\begin{table*}[!h]
\caption{Computed radial velocities $RV$ for HD~188088. Gaussian parameters (depth, width, center) were left free. In the table the width of the best fit Gaussian ($\sigma)$ is also given for each star and observation. }
\begin{tabular}{ccccccc}
\hline \hline
 Cycle & Slit Angle & $RV_{1}$ &    $\sigma_{RV_{1}}$ & $RV_{2}$ & $\sigma_{RV_{2}}$ & $\Delta V$ \\
            & $(^{o})$  & (\kms)      &   (\kms)                      &(\kms)         &       (\kms)               &  (\kms)  \\
\hline
First   &       0       &        2.778     &       4.785   &    -10.613 &       4.605  &      13.391      \\
           &    0       &       2.771   &       4.765   &       -10.627  &       4.566   &        13.398     \\
Cycle &         90      &       2.895   &       4.748   &       -10.555 &       4.554  &      13.450      \\
        &       180     &       2.720   &       4.740   &       -10.841 &       4.541  &      13.561  \\
         &      270     &       2.882   &       4.716    &      -10.568 &       4.547  &      13.450      \\
         &      360     &       2.943   &       4.768    &  -10.408     &       4.609  &       13.351     \\
         &      0       &       2.973   &       4.805   &   -10. 388    &       4.641  &      13.361      \\ 
\hline
Second &        0        &      2.693    &      5.098    &      -10.623  &       4.960    &      13.316   \\
             &  0        &      2.657    &      5.118    &      -10.654  &       4.963    &      13.313   \\
Cycle    &      90       &      3.113    &      5.146    &      -10.323  &       4.958    &      13.436   \\
             &  180      &      3.036    &      5.058   &       -10.540  &       4.902    &      13.576   \\
             &   270   &    2.840         &      4.983          &          -10.592     &     4.881          &      13.432   \\
             &  360      &      2.945    &      5.052    &      -10.342  &       4.914    &      13.287   \\
             &   0       &      2.896    &      5.126    &      -10.388  &       4.986    &      13.284   \\
\hline
\end{tabular} 
\label{tab:RVS}
\end{table*}

\begin{table}[]
\caption{Derived D and $\theta$ for HD~188088. The upper line refers to the first observation cycle, 
the bottom line to the second cycle.  We note that the results for the first cycle are not corrected for the slit effects. When these are taken into account, 7 milliarcseconds must be added to the separation derived for cycle 1.}
\begin{tabular}{cccccc}
\hline \hline
$\Delta V_1$&$\Delta V_2$& $\Delta V_3$ & $\Delta V_4$ & $D$&$\theta$ \\
$($\kms$)$ & $($\kms$)$ & $($\kms$)$ & $($\kms$)$ & (arcsec) & $(^{o})$  \\
\hline
13.376  &      13.561  &     13.450  &     13.450 &     0.012    &      0  \\
13.300  &  13.576   &        13.436  &   13.432  &     0.018    &  -0.4 \\

\hline
\end{tabular}
\label{tab:results}
\end{table}

The results are very consistent and provide a robust distance of 18 $\pm$2 milli-arcseconds between  the stars and a  separation angle of -0.4$\pm$ 5 degrees. In this first application a precision of a few milli-arcseconds is obtained, which can be considered very satisfactory for a first test and shows the potentiality of the method. 

\section{Discussion}

Applying  tachoastrometry to double line spectroscopic binaries enables us to  simultaneously obtain the radial velocity curve and  the  geometry of the systems, providing the separation and the angle between the stars.   In this way it will be possible to determine full orbits for SB2 stars, without the need of interferometric observations. SB2 full orbital solutions  provide accurate masses and distances (see, e.g., \citet{Pourbaix1998}). 
With tachostrometry  distances and masses of SB2 up to several hundred  parsecs could be determined,  providing  access to special classes of objects, such as young stars  and  nearby clusters and associations. 
The concept of SB2 in this context should be considered in a wide sense because it will be  also possible  to separate stars in distant regions that  appear unresolved because of the limited spatial resolution of the telescope, and it is not required that the stars belong to a physical binary system. 

It is worth mentioning that, to the best of our knowledge, the geometrical component has been so far ignored when solving the radial velocity curves  of SB2. The geometrical effect can introduce systematic effects in the K$_1$ and K$_2$ estimates, or can simply add noise to the measurements, depending on the slit orientation used. The effect is clearly small ( $\sim$300 \ms for the star we observed), but might not be negligible when very accurate measurements are needed. It is comparable to or larger than the measurement errors of the radial velocities. 

Even if  we used SB2 stars to introduce the concept of tachoastrometry, in principle  
this technique can   be applied to all systems showing composite spectra. In all astrophysical situations  
where resolved spectral lines are produced in physically distinct regions, tachoastrometry 
can be applied.  Interesting  examples could be stars surrounded by  asymmetric disks,  binaries with low mass companions, 
novae, supernovae, or quasars. In all these cases,  the
emission and absorption line systems (or continuum) are well separated  and may originate in physically distinct places. 
It might  not be trivial to use the technique, but there is no reason why it should not apply. 
Some of the over mentioned sources have rather broad lines, up to many \kms, so  the use of Doppler shift measurements at the few ms$^{-1}$ precision could be questioned. Indeed, obtaining accurate velocity measurements for broad lines is impossible, 
but it is important to note that in tachoastrometry the angular resolution is given by the 
ratio between the measured velocity shift and the resolving power. So, the resolving power of the spectrograph should 
match the line width of the observed objects. In the case of spectral lines thousands of  \kms wide,  one can aim to measure shifts with only a few kms$^{-1}$ precision.  In this case a low resolution spectrograph will be used. What actually determines   the angular resolution is only the ratio between the Doppler velocity precision and the spectrograph scale at the slit (expressed in kms$^{-1}$). So in  the case of broad lines, a spectrograph with a resolving power of 300 for 1 arcsecond slit will still produce  1 milli-arcsecond resolution if  a velocity  precision  of 1 \kms is obtained. 

Even single stars could be  sources of observations, if we consider that 
 different lines can be  formed in spatially distinct structures. The best known are probably the core of the deep absorption lines, 
 such as Ca II H and K,  which  are formed in the stellar chromosphere. These lines are enhanced in active regions, and vary with the stellar rotation period 
 because  of the inhomogeneities on the stellar surface.  Applying tachoastrometry differentially to chromospheric and photospheric lines would provide the position of the active region with respect to the stellar center. 
 Similarly, lines that are very temperature sensitive  strongly react to the presence of cool spots and modulate their intensities 
thanks to the appearance and rotation of strong spots. Doppler imaging has been developed over the years to interpret  these variations  and to produce stellar maps.  The proposed technique can in principle measure the angle of the rotating inhomogeneities  on a stellar disk, even for stars that do not rotate at a fast rate.

We should finally discuss  tachoastrometry with respect to spectroastrometry. They originate from the same work  \citep{Beckers1983}  although they are substantially different in their applications.  Tachoastrometry does not bring more information than spectroastrometry, but  it  is extremely simple to use. It has a more restricted range of applications; for instance, it cannot be applied to systems only with continuum because it is sensitive to spectral lines. In principle, both techniques are  limited by the  S/N ratio and only application to real objects and experience will reveal  their real sensitivity limits. 
We finally note that  tachoastrometry does not depend on the telescope size or on  site quality; it allows small telescopes to reach a very good spatial resolution  in sites that are not optimal  and without the support of adaptive optics. 
 
\section{A simple device to  simultaneously record opposite angles}

\begin{figure}[h]
\includegraphics[scale=0.3]{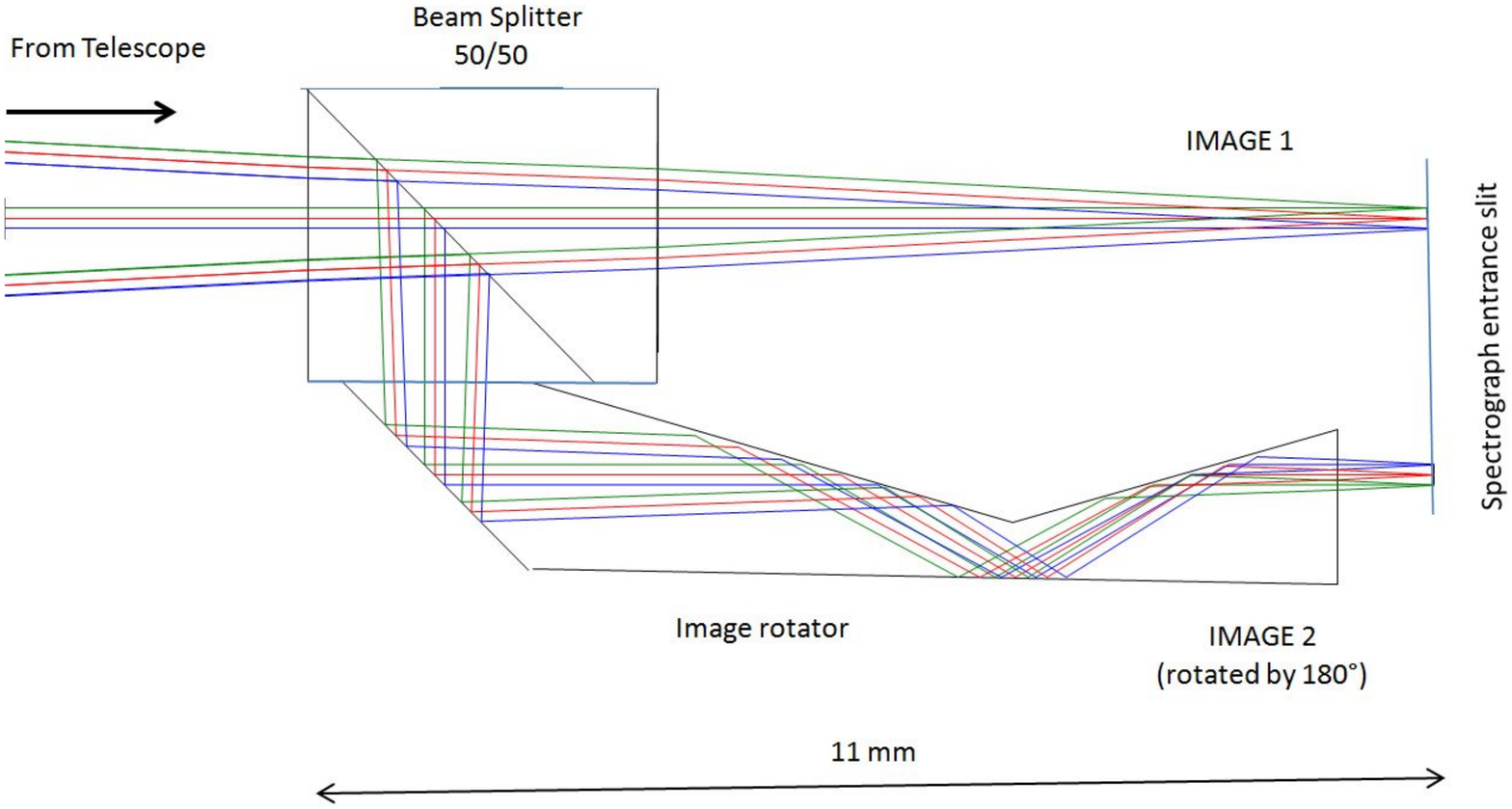}
\caption{Optical design of a simple device to  simultaneously record spectra with 180 degree orientation. The design is optimized for the UVES spectrograph. The rays with different colors do not represent different wavelengths, rather the geometrical position of the extremes, to show visually how  the green ray, which is located at the upper limit at the entrance, and the blue ray, which is located at the lower extreme, maintain the geometry in slit 1 while they are geometrically rotated in slit 2, after an odd number of reflections.   }
\label{fig4}
\end{figure}

It is possible to design a simple device that allows the simultaneous acquisition of two spectra taken with opposite (180 degrees) slit orientation  angles. Such a device  would  improve the quality of the observations and optimize the data reduction, because the
two spectra would be acquired simultaneously and with exactly the same 
instrumental configuration.
The device should be able to rotate the slit by 180 degrees, keeping the two images  co-focal.  The simplest way to rotate an image is to use an odd number of reflections. Figure \ref{fig4} shows a possible device, adapted for the UVES spectrograph. The device is located before the spectrograph slit, in front of the slit plane and the non-dispersed light, coming from the telescope on the left side, is first divided into two halves by a beamsplitter. Half of them pass through a system of five reflections while the others are  not deviated, but  pass through a  glass layer that compensates 
 for the shorter optical path, so that the two paths have the same focal distance. 

 \section {Appendix:  computation of the photon center shift for seeing-limited,  finite slit observations }
 
 In this Appendix we provide formulae and figures to compute the shift of the light center as seen by a spectrograph after a slit. We have computed the  shift for the case of a  seeing-limited,  point-like (Gaussian) source.   The entrance image is produced by the product of the seeing disk and the spectrograph slit.   
The observed shift is determined by the center of light of the photons, i.e., the average light position, for a Gaussian distribution centered at $x_0$ and truncated outside the interval $[-a, a]$ representing the slit aperture.  This is expressed by coupling 

%
%

$$ f(x)   \propto \exp \left[ - \frac{(x - x_0)^2}{2 \sigma^2} \right] $$ 

with 

$$  \langle x \rangle = \frac{\int_{-a}^a x\, f(x) \, dx}{\int_{-a}^a f(x) \, dx}. $$


Called the light center (which is the shift observed after the slit)  $\langle x \rangle$, this is given by 

\begin{equation}
\langle x \rangle = x_0 + \frac{\sqrt{2} \sigma \left[ \exp \left( -\frac{(x_0 + a)^2}{2 \sigma^2} \right) - \exp \left( -\frac{(x_0 - a)^2}{2 \sigma^2} \right) \right]}{\sqrt{\pi} \left[ \erf \left( \frac{x_0 + a}{\sqrt{2} \sigma} \right) - \erf \left( \frac{x_0 - a}{\sqrt{2} \sigma} \right) \right]}
\label{bary}
\end{equation}

Figure \ref{fig5} shows the computation of the observed barycenter  for different seeing values, varying from 0.5 to two arcseconds,  and with a slit width fixed to  one arcsecond.  This figure provides  the general trends for the shifts  because the only relevant quantity is the ratio between slit width and seeing. So any realistic combination can be extrapolated from this figure.  
The computed shifts are extreme, extending them to values that are larger than the slit width, which is represented  by the red vertical lines in the figure. 
When the PSF is much smaller than the slit width the observed shift is, as expected, similar to the real source shift, for reasonable small values. 
When the PSF is larger than the slit width, a shift is still present and can be shown that it is given by: 

\begin{figure*}[h]
\includegraphics[]{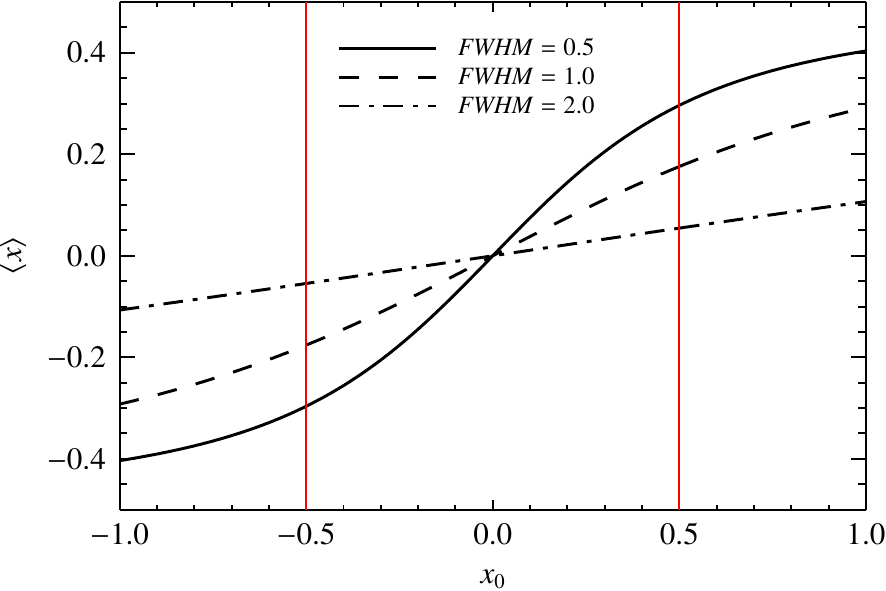}
\caption{Observed barycenter shift as a function of real shift of the star in the focal plane, in arcseconds. Vertical red lines show the slit limits. The slit width is one arcsec.  Three examples are shown: seeing much better than  (0.5 arcsec), comparable to (1 arcsec), and larger  than (2 arcsec) slit width. }
\label{fig5}
\end{figure*}

\begin{equation}
\langle x \rangle = x_0 \frac{a^2}{3 \sigma^2} \
\end{equation}

Equation (12)  is actually usable as a first approximation (slightly overestimating) to compute the shifts even  for those intermediate cases in which the PSF is comparable with the slit width. As an example, for 1 arcsecond seeing and 1 arcsecond slit,   it would predict  $\Delta x_{obs}$ = 0.23 arcseconds for a $\Delta x_{sky}$= 0.5 arcseconds, while the correct formula would predict 0.18 arcseconds.   It can be used therefore for a quick, rough estimate.


\begin{acknowledgements}
LP thanks J. Beckers for   enlightening discussions  while observing many years ago at the ESO 3.6m telescope. He also  acknowledges the Visiting Researcher program of the CNPq Brazilian Agency, at the Fed. Univ. of Rio Grande do Norte, Brazil. Support for CC is provided by the Ministry for the Economy, Development, and Tourism's Programa Iniciativa Cient\'{i}fica Milenio through grant IC\,120009, awarded to the Millennium Institute of Astrophysics (MAS).
\end{acknowledgements}

\bibliographystyle{aa}
\bibliography{ccortesabib}
%
%
\end{document}